\documentclass[aps,twocolumn,amsmath,amssymb,reprint,numbers,superscriptaddress,noeprint,longbibliography]{revtex4-1}

\usepackage{makecell}
\usepackage{soul}
\usepackage{geometry}
\usepackage{graphicx}
\usepackage{booktabs}
\usepackage{array}
\usepackage{relsize}
\usepackage{color}
\usepackage{xcolor}
\usepackage{eqnarray}
\usepackage{bm}
\usepackage{caption}
\usepackage{textcase}
\usepackage{textcomp}
\usepackage{braket}
\usepackage{setspace}
\usepackage{mathtools}
\usepackage{setspace}

\usepackage{subfig}
\usepackage{tikz}
\usetikzlibrary{positioning}
\usepackage{hyperref}
\hypersetup{
     colorlinks   = true,
     linkcolor    = blue,
     citecolor    = blue,
     urlcolor = blue
}

\definecolor{cola}{rgb}{0.7,0.1,0.1}
\definecolor{colb}{rgb}{0.8,0.3,0}
\definecolor{colc}{rgb}{0.3,0.7,0}
\definecolor{cold}{rgb}{0,0.35,0.75}
\definecolor{cole}{rgb}{0.63, 0.13, 0.94}

\definecolor{darkerred}{rgb}{0.8,0,0}
\usepackage[normalem]{ulem}

\newcommand{\aver}[1]{\left \langle #1 \right \rangle}

\renewcommand{\phi}{\varphi}
\renewcommand{\kappa}{\varkappa}

\renewcommand{\braket}[3]{\left \langle #1 \right. \left | #2 \right | \left.  #3 \right \rangle}

\begin{document}

\title[]{Optical spin control and coherence properties of acceptor bound holes in strained GaAs}

\author{Xiayu~Linpeng}
\affiliation{Department of Physics, University of Washington, Seattle, Washington 98195, USA}
\author{Todd~Karin}
\affiliation{Lawrence Berkeley National Laboratory, Berkeley, CA 94720, USA}
\author{Mikhail V. Durnev}
\affiliation{Ioffe Institute, 194021 St.-Petersburg, Russia}
\author{Mikhail M. Glazov}
\affiliation{Ioffe Institute, 194021 St.-Petersburg, Russia} 
\affiliation{Spin Optics Laboratory, Saint Petersburg State University, 198504 St. Petersburg, Russia}
\author{R\"udiger Schott}
\affiliation{Lehrstuhl f\"ur Angewandte Festk\"orperphysik, Ruhr-Universit\"at Bochum, D-44870 Bochum, Germany}
\author{Andreas D. Wieck}
\affiliation{Lehrstuhl f\"ur Angewandte Festk\"orperphysik, Ruhr-Universit\"at Bochum, D-44870 Bochum, Germany}
\author{Arne Ludwig}
\affiliation{Lehrstuhl f\"ur Angewandte Festk\"orperphysik, Ruhr-Universit\"at Bochum, D-44870 Bochum, Germany}
\author{Kai-Mei~C.~Fu}
\affiliation{Department of Physics, University of Washington, Seattle, Washington 98195, USA}
\affiliation{Department of Electrical Engineering, University of Washington, Seattle, Washington 98195, USA}

\begin{abstract}

Hole spins in semiconductors are a potential qubit alternative to electron spins. In nuclear-spin-rich host crystals like GaAs, the hyperfine interaction of hole spins with nuclei is considerably weaker than that for electrons, leading to potentially longer coherence times. Here we demonstrate optical pumping and coherent population trapping for acceptor-bound holes in a strained GaAs epitaxial layer. We find \textmu s-scale longitudinal spin relaxation time T$_1$ and an inhomogeneous dephasing time T$_2^*$ of $\sim$7~ns. We attribute the spin relaxation mechanism to a combination effect of a hole-phonon interaction through the deformation potentials and a heavy-hole light-hole mixing in an in-plane magnetic field. We attribute the short T$_2^*$ to g-factor broadening due to strain inhomogeneity. T$_1$ and T$_2^*$ are quantitatively calculated based on these mechanisms and compared with the experimental results. While the hyperfine-mediated decoherence is mitigated, our results highlight the important contribution of strain to relaxation and dephasing of acceptor-bound hole spins.

\end{abstract}

\date{\today}

\maketitle


\section{Introduction}

Spin systems in semiconductors have been actively studied due to the potential applications for nanoscale spintronics and quantum information technologies. Significant effort has been focused on electron spins in low-dimensional systems, e.g. quantum dots and donors~\cite{ref:Kloeffel2013psb, ref:Morello2020dss, ref:Linpeng2018cps}. However, due to the hyperfine interaction with the nuclei in the host crystal, the coherence time of electron spins can be short, typically on the nanosecond scale. Isotopic purification can significantly reduce this effect in group-IV and group II-VI semiconductors, e.g. in diamond and silicon. For group III-V semiconductors such as GaAs, this technique is not applicable as there is no stable isotope with zero nuclear spins. An alternative solution is to use hole spins which have a much weaker hyperfine interaction due to the $p$-symmetry of the hole Bloch wave function~\cite{glazov2018electron}. Research in III-V quantum dots has shown \textmu s-scale hole-spin coherence times~\cite{ref:Brunner2009csh,ref:Prechtel2016dhs}, compared to ns-scale in electron spins~\cite{ref:Xu2008cpt,ref:Fu2005cpt}. Spin control techniques such as optical pumping, coherent population trapping (CPT) and ultra-fast optical control have been demonstrated~\cite{ref:Brunner2009csh,ref:Prechtel2016dhs,ref:Greve2011ucc}. Remote entanglement between two hole spins has been performed leveraging this enhanced coherence time~\cite{ref:Delteil2015ghe}. In addition to the enhanced coherence properties, faster electronic gate operations are possible due to large spin-orbit interaction, as demonstrated in silicon and germanium quantum dots~\cite{ref:Watzinger2018ghs,ref:Maurand2016css}. 

A hole bound to an acceptor is an analogous spin qubit system to a hole-doped quantum dot with the added feature of high optical homogeneity~\cite{ref:Karin2015rpm}. However, because of the degeneracy of the heavy hole (hh) and light hole (lh) valence bands of GaAs, the strong spin-orbit interaction results in a short spin relaxation time, typically much less than 1 ns~\cite{ref:Hilton2002oof,ref:Vina1992srd}. This is not a problem in quantum dots as the mixing between hh and lh is significantly suppressed by the large hh-lh splitting due to strain and spatial confinement~\cite{ref:Heiss2007oes}. By analogy, if a large strain can be introduced to a p-doped GaAs crystal, relaxation times much longer than ns should also be achievable for hole spins bound to acceptors.


In this paper, we apply 0.04\% compressive strain to a GaAs epitaxial layer and study the optical and spin properties of an ensemble of acceptors. We demonstrate optical pumping and CPT for the acceptor system in this strained GaAs sample. Microsecond-scale longitudinal hole spin relaxation time T$_1$ is observed. The measured magnetic-field dependence of T$_1$ can be explained by a combination effect of a hole-phonon interaction through the deformation potentials and a hh-lh mixing due to an in-plane magnetic field. A $\sim$7~ns decoherence time T$_2^*$ is extracted from the CPT measurements. This time is much shorter than the measured $>$100~ns T$_2^*$ for single hole spins in III-V quantum dots, determined by similar CPT measurements~\cite{ref:Brunner2009csh,ref:Prechtel2016dhs}. We attribute the short T$_2^*$ to hole g-factor broadening due to strain inhomogeneity in the ensemble. We theoretically calculate the intrinsic T$_2^*$ from the hyperfine interaction with nuclei to be 58~ns.

The paper is organized as follows. Section~\ref{sec:sample} gives a brief description of the strained sample and the experimental setup. In Sec.~\ref{sec:strainandPL}, we analyze the photoluminescence (PL) properties from acceptors and show how we calculate the strain from the PL spectra. In Sec.~\ref{sec:T1exp} and Sec.~\ref{sec:CPT}, we show the measurement techniques and the measured results for T$_1$ and T$_2^*$. The mechanisms of T$_1$ and T$_2^*$ are briefly discussed in these two sections. In Sec.~\ref{sec:T1theory}, we present detailed theory on the calculation of T$_1$. The paper ends with a brief conclusion in Sec.~\ref{sec:conclusion}.

\section{Sample description and experimental setup}
\label{sec:sample}

\begin{figure*}[t]
  \centering
  \includegraphics[width=7 in]{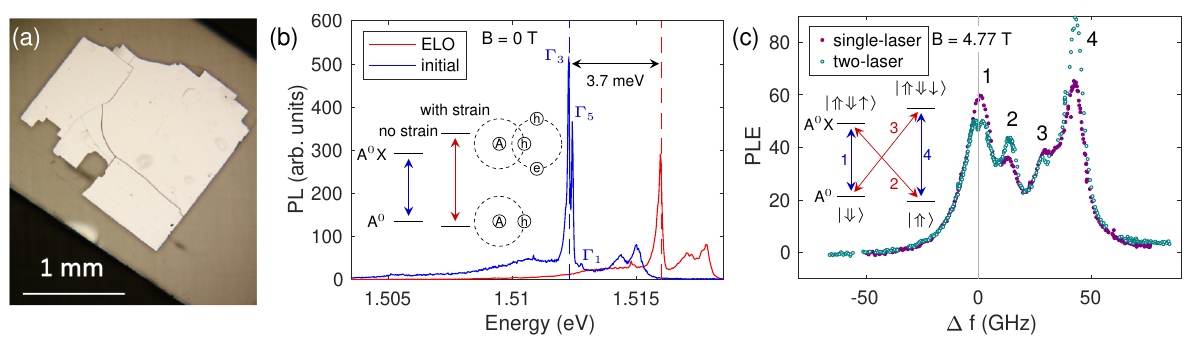}
  \caption{\label{fig:sample} (a) Optical microscope image of the a GaAs epitaxial layer transferred to the MgO substrate. (b) PL spectra of the GaAs epitaxial layer before and after ELO process at 1.5~K and 0~T. Excitation at 1.653~eV with 80~nW power. The laser spot diameter is $\sim$1~\textmu m. The inset shows the cartoon model of the acceptor systems and how the energies of $\text{A}^0$ and $\text{A}^0\text{X}$ change with strain. In the cartoon diagram, ``A" denotes the acceptor center, ``h" denotes hole, and ``e" denotes electron.  (c) Single-laser and two-laser PLE spectra at 1.5~K and 4.77~T. The single-laser PLE spectrum is taken by scanning a laser across all four transitions and collecting the signal from THT. A typical THT spectrum is shown in Appendix~\ref{app:THT}. The two-laser PLE spectrum is taken with a second laser fixed at the energy of transition 1. $\Delta\text{f}$ is the detuning of the scanning laser with respect to the energy of transition 1. We have used background subtraction on both the single-laser and two-laser PLE spectra where we use the PLE intensity at large $\Delta\text{f}$ as the background. All lasers are at 1~\textmu W and 45 degree polarization. The laser spot diameter is $\sim$1~\textmu m. The inset shows the energy structure of the acceptor system. Transitions 1, 4 (2, 3) are polarized in the horizontal (vertical) direction.  
  }
\end{figure*}

The strained sample consists of a 2~\textmu m (001) $p$-type GaAs epitaxial layer on a MgO substrate. The GaAs layer is doped with carbon with an acceptor density of ${\sim}2.5{\times}10^{14}$~cm$^{-3}$, determined from Hall measurements. The GaAs is transferred and bonded to the MgO substrate through an epitaxial lift-off (ELO) process at room temperature (see Appendix~\ref{app:ELO}). Compressive strain is introduced to GaAs when the sample is cooled down to 1.5~K due to the different thermal expansion rate of GaAs and MgO. The MgO substrate is chosen as the carrier as it is transparent at the band gap of GaAs and can produce significant compressive strain. An optical microscope image of the transferred GaAs epitaxial layer on MgO is shown in Fig.~\ref{fig:sample}(a). We note that this ELO method is not ideal; some cracking is observed, and possible slippage between the membrane and substrate can result in both strain reduction and strain inhomogeneity.

The photoluminescence (PL) of the sample is studied using a home-built confocal microscope with a resolution of $\sim$1~\textmu m. The sample is cooled to 1.5~K in a helium-immersion magnetic cryostat (Janis SOM). The hole spin states are controlled and measured with two tunable continuous-wave Ti:Sapphire lasers (Spectra-Physics Matisse and Coherent 899-21). In pulsed experiments, the laser pulse is generated by passing the laser through an acousto-optic modulator (Gooch\&Housego) with an on/off extinction ratio $>10^4$.

\section{Induced strain and Photoluminescence properties}
\label{sec:strainandPL}

\begin{table}[!bp]
\begin{center}
\begin{tabular}{c c c c c c c}

\Xhline{2\arrayrulewidth}
$a_c$ (eV) & $a_v$ (eV) & $C_{12}/C_{11}$ & $b$ (eV)  \\
\Xhline{1\arrayrulewidth}   
-7.17 & 1.16  & 0.4526 & -1.7 \\
\Xhline{2\arrayrulewidth}
\end{tabular}
\end{center}
\caption{Parameters used to calculate the strain and energy shift in ELO~\cite{ref:Chuang2009ppd}.}
\label{table:constant}
\end{table}

Figure~\ref{fig:sample}(b) shows the PL spectra before and after ELO at 0~T and 1.5~K. The main sharp peaks are from the transitions between the acceptor bound exciton  (A$^0$X) and the neutral acceptor (A$^0$) states. In the unstrained sample, three acceptor peaks $\Gamma_3$, $\Gamma_5$ and $\Gamma_1$ are observed due to the different hh and lh states for the two holes in A$^0$X, with the energy splitting due to hole-hole coupling and the crystal field~\cite{ref:Karin2015rpm}. As compressive strain introduces a hh-lh splitting and a hh-like ground state, the two holes in A$^0$X will be in the hh spin-singlet state, which corresponds to the single acceptor peak in the PL spectrum of the strained sample. Compared with the unstrained sample, a $\sim$3.7~meV blue shift of the acceptor transition is observed. This energy shift is mainly caused by the shift of conduction and valence bands under strain and can be used as an estimate for the change in band gap energy. The strain is estimated from the energy shift by $\Delta E = 2 (a_c - a_v) (1 - C_{12}/C_{11}) u_{xx}$, where $a_c$ and $a_v$ are the deformation potentials that determine the conduction and valence band shift, and $C_{ij}$ are the components of the elastic stiffness tensor in GaAs, see Table~\ref{table:constant}. From this, we obtain the value of the in-plane strain in our sample $u_{xx} \approx u_{yy} \sim -0.04 \%$. This strain leads to the splitting of the heavy and light holes subbands $E_{hh} - E_{lh} = 2 b (1 + 2 C_{12}/C_{11}) u_{xx}$, where $b$ is the valence-band deformation potential that determines the hh-lh splitting.  We find $E_{hh} - E_{lh} \sim 2.6$~meV for GaAs parameters~\footnote{In these estimates we disregarded the renormalization of the deformation potential due to the Coulomb potential of the acceptor~\cite{Kogan1981}, this effect is discussed in more detail below in relation to the spin-flip of the hole.}. Since the splitting is positive for compressive strain, and is much larger than the 0.13~meV thermal energy at 1.5~K, 
the majority of the holes populate the hh-like state. 



The strained sample is studied in an in-plane magnetic field ($B\bot[001]$). As shown in the inset of Fig.~1(c), there are four allowed optical transitions: transition 1 ($\ket{\Downarrow} \leftrightarrow \ket{\Uparrow\Downarrow\uparrow}$), transition 2 ($\ket{\Uparrow} \leftrightarrow \ket{\Uparrow\Downarrow\uparrow}$), transition 3 ($\ket{\Downarrow} \leftrightarrow \ket{\Uparrow\Downarrow\downarrow}$), and transition 4 ($\ket{\Uparrow} \leftrightarrow \ket{\Uparrow\Downarrow\downarrow}$). Here, $\ket{\uparrow}$ ($\ket{\downarrow}$) and $\ket{\Uparrow}$ ($\ket{\Downarrow}$) denote the eigenstates of the electron and hole in the in-plane field. The splitting between states $\ket{\Uparrow}$ and $\ket{\Downarrow}$ is due to the hole Zeeman splitting. The splitting between states $\ket{\Uparrow\Downarrow\uparrow}$ and $\ket{\Uparrow\Downarrow\downarrow}$ is due to the electron Zeeman splitting, as the two holes are in a spin singlet state. As shown by the PL spectra in Appendix~\ref{sec:gfactor}, transitions 1 and 4 are horizontally polarized (parallel to the magnetic field), while transitions 2 and 3 are vertically polarized (perpendicular to the magnetic field). The selection rules (Appendix~\ref{app:selectionrules}), together with the reasonable assumption  $g_e^\perp < 0$ for the electron $g$-factor, yield $g_{hh}^\bot < 0$.
The measured electron and hole g-factors are thus $g_e^\bot = -0.43$ and $g_{hh}^\bot = -0.15$.

A single-laser photoluminescence excitation (PLE) spectrum is taken to resolve all four transitions, as shown in Fig.~\ref{fig:sample}(c). The single-laser PLE spectrum is taken by scanning a laser across the A$^0$X transitions and collecting the signal from the two-hole transitions (THT). In order to excite all four transitions, the laser is linearly polarized at 45$^\circ$ with respect to the magnetic field. From the spectrum, the inhomogenous linewidth of the acceptor transitions is $\sim$10~GHz. A two-laser PLE spectrum is taken to confirm the validity of the energy diagram shown in the inset of Fig.~\ref{fig:sample}(c). In the two-laser PLE spectrum, one laser is fixed at transition 1 and a second laser is scanned across all four transitions. Compared to the single-laser PLE, there is a decrease in signals from transitions 1 and 3, and an enhancement of transitions 2 and 4. These changes are consistent with the effect of optical pumping where the fixed laser pumps the spin states from $\ket{\Uparrow}$ to $\ket{\Downarrow}$. Signals from transitions 1 and 3 (2 and 4) decrease (increase) as the intensity of these two transitions are proportional to the population in $\ket{\Uparrow}$ ($\ket{\Downarrow}$). Additionally, a small dip in peak 1 is observed which is attributed to spectral hole burning. A fit of the dip shows a homogeneous linewidth of $\sim$1~GHz, corresponding to a $\sim$0.2~ns A$^0$X radiative lifetime, which is in reasonable agreement with experimental measurements~\cite{ref:Karin2015rpm}. A small dip in peak 2 is expected due to CPT which will be further discussed in Section~\ref{sec:CPT}. However, due to the scan resolution, the CPT dip is not clearly resolved in this spectrum. 

\section{Optical pumping and T$_1$ measurement}
\label{sec:T1exp}

\begin{figure}[t]
  \centering
  \includegraphics[width=3.5 in]{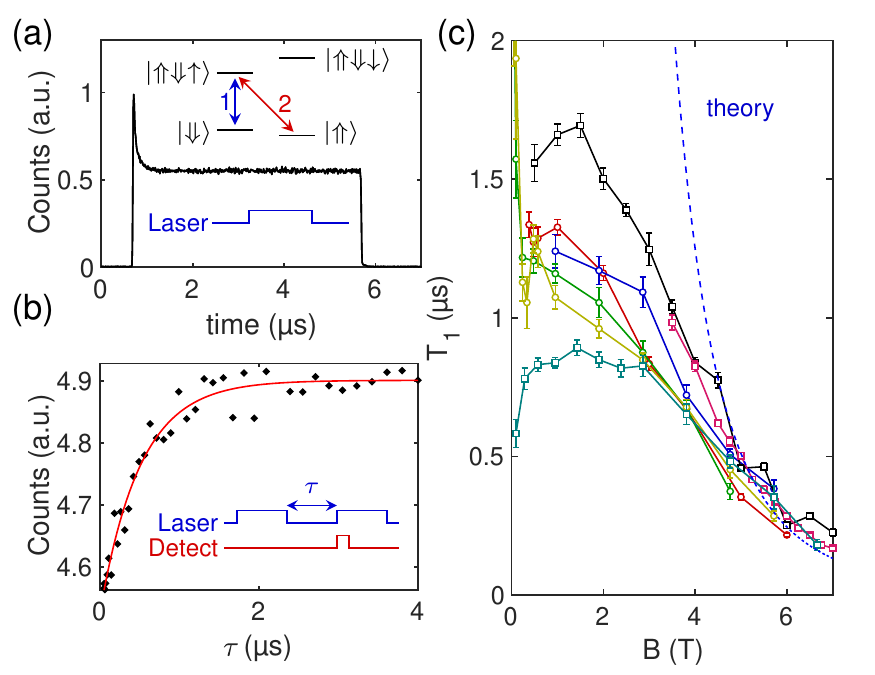}
  \caption{\label{fig:T1} (a) An optical pumping curve at 1.5~K and 1.9~T. The inset shows the laser sequence. The laser is resonantly on transition 1 with 45~nW power and the PL from transition 2 is collected through a single photon counting module. The laser spot diameter is $\sim$1~\textmu m. The insets show the energy diagram and the laser sequence. The detection is on all the time during the laser sequence. (b) A population recovery curve at 1.5~K and 1.9~T. The energy of the laser and detection are the same as {\bf a}. A single exponential curve is used to fit for the T$_1$. T$_1 = 0.51\pm0.04$~\textmu s for this data. The inset shows the laser sequence. The detection window is 0.8~\textmu s. (c) T$_1$ as a function of the magnetic fields. Different colors represent different locations on the sample. The dashed line shows the curve from theoretical calculation, Eq.~\eqref{theory:T1:fin}.
  }
\end{figure}
In the A$^0 \leftrightarrow \text{A}^0$X system, the A$^0$ holes can be initialized to a certain spin state by optical pumping. As shown in Fig.~\ref{fig:T1}(a), a 5~\textmu s laser pulse is applied resonantly on transition 1 so the spin states are pumped from spin $\ket{\Uparrow}$ to spin $\ket{\Downarrow}$. The PL signal from transition 2, which is proportional to the population of $\ket{\Uparrow}$, is recorded during the optical pumping pulse. A decrease of the spin population is observed, indicating partial spin initialization is achieved. 

The spin relaxation time T$_1$ is measured by initializing the spin to $\ket{\Downarrow}$, and measuring the recovery of the PL from transition 2 as a function of variable time $\tau$. A single exponential fit is used to extract T$_1$ from the recovery curve, as shown in Fig.~\ref{fig:T1}(b). T$_1$ as function of magnetic field at different spots on the sample is shown in Fig.~\ref{fig:T1}(c). Between 5 and 7~T, the T$_1$ at different spots is similar, following approximately a B$^{-3}$ dependence. At these high fields, we attribute the spin relaxation of holes to the admixture mechanism resulting from the hh-lh mixing by magnetic field and a hole-phonon interaction through the deformation potentials. The detailed theory is discussed in Sec.~\ref{sec:T1theory}. The calculation based on this theory matches the experimental result, as shown in Fig.~\ref{fig:T1}(c). T$_1$ longer than 100~\textmu s  has been measured in self-assembled InGaAs quantum dots due to the much larger hh-lh splittings~\cite{ref:Heiss2007oes}. Such long T$_1$ can potentially also be achieved in the acceptor system by applying stronger and more homogeneous strain, which could be realized with other strain engineering techniques such as wafer bonding~\cite{ref:Stanton2020esh} or using piezoelectric actuators~\cite{ref:Yuan2018usf}.

Below 5~T, T$_1$ is noticeably different at different locations on the sample and does not have a clear B-field dependence. A possible mechanism to explain this behavior is a combination of a hole-hole exchange interaction and inhomogeneous hyperfine fields, which is shown to be a mechanism for T$_1$ of donors at low fields~\cite{ref:Linpeng2016lsr}. This interaction depends on the local environment, which can vary across the sample.

\section{Coherent population trapping and the spin decoherence time T$_2^*$}\label{sec:CPT}

\begin{figure}[tb]
  \centering
  \includegraphics[width=3.2 in]{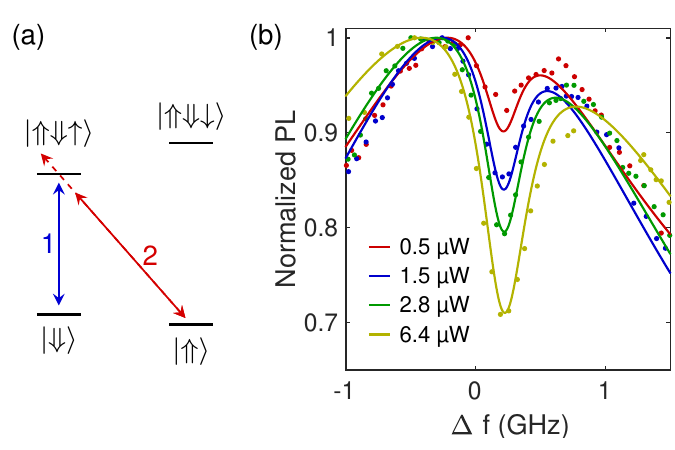}
  \caption{\label{fig:CPT} (a) Energy diagram of the CPT experiment. The control laser is fixed at transition 1 and the probe laser is scanned across transition 2. (b) CPT with different probe laser power. Each curve is a two-laser PLE spectrum where $\Delta\text{f}$ is the detuning of the probe laser compared to the energy of transition 2. The solid curves are from a simultaneous fit of the data at all different probe laser powers using the 3-level density matrix model. The control laser is fixed at transition 1 with a slight detuning of about 0.2 GHz and with a power of 3~\textmu W. The polarization of both lasers are set at 45 degrees. The laser spot diameter is $\sim$1~\textmu m. The temperature is 1.5~K and the magnetic field is 7~T. We note that we have used background subtraction on all CPT curves where we use the signal at large $\Delta\text{f}$ as the background.
  }
\end{figure}

Next, we perform coherent population trapping (CPT) on the A$^0$-A$^0$X system to investigate the hole-spin coherence properties. As shown in Fig.~\ref{fig:CPT}(a), the A$^0$X state $\ket{\Uparrow\Downarrow\uparrow}$, together with the two A$^0$ states, $\ket{\Uparrow}$ and $\ket{\Downarrow}$, form a $\Lambda$-system. With a control laser driving the transition 1 ($\ket{\Uparrow}\leftrightarrow\ket{\Uparrow\Downarrow\uparrow}$) and a probe laser driving the transition 2 ($\ket{\Downarrow}\leftrightarrow\ket{\Uparrow\Downarrow\uparrow}$), a destructive interference occurs when the energy difference between the two lasers equals the hole Zeeman splitting. On two-laser resonance, the system is pumped into a dark state, i.e. a superposition state between $\ket{\Uparrow}$ and $\ket{\Downarrow}$~\cite{ref:Harris1997eit, Agap_ev_1993}.

In our experiment, CPT is revealed by the two-laser PLE spectrum. The energy of the control laser is fixed near the resonance of transition 1 and the probe laser is scanned across the transition 2. A dip in the PLE spectrum occurs when the probe laser is on resonance with transition 2, as shown in Fig.~\ref{fig:CPT}(b). The linewidth and depth of the dip depend on the laser powers, the spontaneous emission rate of the $\ket{\Uparrow\Downarrow\uparrow}$ state, and the dephasing rate between $\ket{\Uparrow}$ and $\ket{\Downarrow}$. The CPT phenomenon is simulated by solving the master equation of a 3-level density matrix considering all relaxation and dephasing terms (see Appendix~\ref{app:dmCPT}).

In Fig.~\ref{fig:CPT}, the CPT curves at different probe powers are simultaneously fit with the density matrix model. The inhomogeneous dephasing time T$_2^*$ between the two spin states is found to be $\sim$7~ns. The measured T$_2^*$ in our sample is significantly smaller than the $>$100~ns T$_2^*$ measured in single III-V quantum dots~\cite{ref:Brunner2009csh,ref:Prechtel2016dhs}. We attribute the short T$_2^*$ to g-factor broadening due to strain inhomogeneity in the sample. The 7~ns T$_2^*$ corresponds to a 0.2\% broadening of the in-plane hole g-factor $g_{hh}^\perp$. This g-factor broadening introduces inhomogeneity to the spin splitting between $\ket{\Uparrow}$ and $\ket{\Downarrow}$, and hence to the spin procession frequency, which leads to a stronger dephasing. In Appendix~\ref{sec:T2theory} we calculate the intrinsic spin dephasing time due to dipole-dipole hyperfine interaction to be $T_2^* \approx 58$~ns. This longer coherence time could be achieved by using strain engineering techniques that provide more homogeneous strain.

\section{Theory of the heavy-hole longitudinal spin relaxation time T$_1$}
\label{sec:T1theory}

\subsection{Wave functions of acceptor-bound holes}

The acceptor-bound hole states in cubic semiconductors are determined by an interplay of the spin-orbit interaction and the Coulomb energy. Before calculating the spin relaxation times, we establish the form of the hole wavefunctions and Zeeman effect in the studied system where strain is present.

In the spherical approximation, the hole bound to an acceptor is described by the total angular momentum $\bm F$, which is the sum of the free-hole angular momentum $\bm J$ ($J = 3/2$ originating from the valence band Bloch functions) and the orbital momentum $\bm L$ of the hole moving in the Coulomb field of an acceptor. In the ground state of the hole, $F = 3/2$ and $L = 0, \,2$~\cite{ref:Baldereschi1973sms,gelmont-dyakonov}. In the absence of external fields, this state is four-fold degenerate with respect to the projection of the total angular momentum $F_z$. The corresponding wave functions $\ket{F_z}$ are~\cite{ref:Baldereschi1973sms,gelmont-dyakonov}
%
\begin{equation}
\begin{aligned}
&\ket{\pm 3/2} = \left[ f(r) Y_0^0(\theta,\varphi) + \frac{g(r)}{\sqrt{5}} Y_2^0(\theta,\varphi) \right] \ket{J, \pm 3/2} \\
& \,\,\, - \sqrt{\frac{2}{5}} g(r) \left[ Y_2^{{\pm}1}(\theta,\varphi) \ket{J,\pm1/2} - Y_2^{{\pm}2}(\theta,\varphi) \ket{J, \mp 1/2} \right]\:, \\
& \ket{\pm 1/2} = \left[ f(r) Y_0^0(\theta,\varphi) - \frac{g(r)}{\sqrt{5}} Y_2^0(\theta,\varphi) \right] \ket{J,\pm 1/2} \\
& \,\,\, +\sqrt{\frac{2}{5}} g(r) \left[ Y_2^{\mp{1}}(\theta,\varphi) \ket{J,\pm 3/2} + Y_2^{\pm2}(\theta,\varphi) \ket{J, \mp 3/2} \right] ,\\
\end{aligned}
\label{eq:newbasis}
\end{equation}
where $Y_l^m(\theta,\phi)$ are the spherical harmonic functions, $f(r)$ and $g(r)$ are the radial parts of the envelope functions, and $\ket{J,J_z}$ are the Bloch functions of the $\Gamma_8$ band. Throughout Sec.~\ref{sec:T1theory} we use the hole representation for the hole wave functions and energies in contrast to the discussion of selection rules and energy diagrams where we used the electron representation. Note, however, that the sign of the hole $g$-factor, as well as the final answer for $T_1$, is independent of the representation used. 


Now let us consider the effect of strain and magnetic field on the ground acceptor state. The biaxial strain induces the splitting between $\ket{\pm 3/2}$ and $\ket{\pm 1/2}$ states discussed in Sec.~\ref{sec:strainandPL}. Since the estimated value of this splitting (${\sim2.6}$~meV) is much smaller than the hole binding energy (${\sim25}$~meV)~\cite{ref:Baldereschi1973sms, ref:Karin2015rpm}, it is possible to neglect the coupling of the ground acceptor state to the excited ones and consider the quadruplet~\eqref{eq:newbasis} only.  In the presence of strain and magnetic field $\bm B \parallel x$, 
the Hamiltonian describing the ground state of an acceptor-bound hole in the basis (Eq.~\ref{eq:newbasis}) reads
\begin{equation}
\label{eq:Ham0}
    \mathcal H_0 = -\frac{\Delta_0}{2} F_z^2 + g_0 \mu_B B F_x\:.
\end{equation}
Here $\Delta_0 > 0$ is the strain-induced splitting between $\ket {\pm 3/2}$ and $\ket{\pm 1/2}$ doublets, $g_0$ is the $g$-factor of an acceptor-bound hole in the absence of strain (accounting for the Coulomb effects)~\cite{gelmont1973, ref:Malyshev1997mma}, $F_j$ are the matrices of the angular momentum $F = 3/2$, and $\mu_B$ is the Bohr magneton.

The Hamiltonian~\eqref{eq:Ham0} does not yield linear in $B$ splitting of $\ket{\pm 3/2}$ doublet, observed in the experiment, see Sec.~\ref{sec:strainandPL}, and is insufficient to describe the experimental data. A nonzero heavy-hole in-plane $g$-factor $g_{hh}^\perp$ results from cubic symmetry terms $\sum_\alpha F_\alpha^3 B_\alpha$ allowed in zinc-blende semiconductors, however these terms are small, since they originate from the coupling with remote electronic bands~\cite{marie1999}. Larger values of $g_{hh}^\perp$ might result from the presence of anisotropic in-plane strain, i.e. nonzero $u_{xy}$ or $u_{xx} - u_{yy}$ components of the strain tensor, in our sample.
This anisotropic strain might be attributed to imperfect bonding between the GaAs epitaxial layer and the MgO substrate. The experimentally observed selection rules are consistent with $|u_{xx} - u_{yy}| \gg |u_{xy}|$ (Appendix~\ref{app:selectionrules}), and therefore we consider the case $u_{xx} \neq u_{yy}$, $u_{xy} = 0$ in the following.
In this case, the hole Hamiltonian has the form
\begin{equation}
    \label{eq:Ham_shear}
    \mathcal H = \mathcal H_0 + \frac{\Delta_1}{2} \left( F_x^2 - F_y^2 \right) \:,
\end{equation}
where the additional term $\propto \Delta_1 \propto u_{xx} - u_{yy}$ accounts for the anisotropic in-plane strain in the sample.

The Hamiltonian~\eqref{eq:Ham_shear} couples the $\ket{\pm 3/2}$ states resulting in the splitting of the doublet. As a result, the new states $\ket{\Uparrow}$ and $\ket{\Downarrow}$ with energies $\varepsilon_{\Uparrow, \Downarrow} = -9\Delta_0/8 \pm g_{hh}^\perp \mu_B B/2$ are formed. At $|\Delta_1|, |g_0 \mu_B B| \ll \Delta_0$, relevant to experimental conditions, the corresponding transverse $g$-factor and wave functions of the heavy-hole states are
\begin{equation}
    \label{g_inplane}
    g_{hh}^\perp = -\frac{3 \Delta_1}{\Delta_0} g_0\:,
\end{equation}
\begin{eqnarray}
    \label{eq:wfs_mixed}
    \ket{\Uparrow} = \frac{1}{\sqrt{2}} \left[ \ket{+3/2} - \frac{\sqrt{3}}{2 \Delta_0} (\Delta_1 + g_0 \mu_B B) \ket{+1/2} \right. \\ 
    \left. - \frac{\sqrt{3}}{2 \Delta_0} (\Delta_1 + g_0 \mu_B B) \ket{-1/2} + \ket{-3/2} \right] \nonumber\:, \\
    \ket{\Downarrow} = \frac{1}{\sqrt{2}} \left[ \ket{+3/2} + \frac{\sqrt{3}}{2 \Delta_0} (\Delta_1 - g_0 \mu_B B) \ket{+1/2} \right. \nonumber \\ 
    \left. - \frac{\sqrt{3}}{2 \Delta_0} (\Delta_1 - g_0 \mu_B B) \ket{-1/2} - \ket{-3/2} \right]  \nonumber\:.
\end{eqnarray}
We stress that in Eqs.~\eqref{g_inplane} and \eqref{eq:wfs_mixed} the Zeeman energy $g_0 \mu_B B$ and the strain-induced coupling parameter $\Delta_1$ can be comparable in magnitude.

It follows from Eq.~\eqref{eq:wfs_mixed} that optical 
transitions $\ket{\Downarrow} \leftrightarrow \ket{\Uparrow\Downarrow\uparrow}$ and $\ket{\Uparrow} \leftrightarrow \ket{\Uparrow\Downarrow\downarrow}$ are 
active in $x$-polarization, whereas transitions $\ket{\Uparrow} \leftrightarrow \ket{\Uparrow\Downarrow\uparrow}$ and $\ket{\Downarrow} \leftrightarrow \ket{\Uparrow\Downarrow\downarrow}$ are 
active in $y$-polarization. By comparison with 
Fig.~\ref{fig:sample}c we conclude that 
$\varepsilon_{\Downarrow} > \varepsilon_{\Uparrow}$, and 
hence, $g_{hh}^\perp < 0$.

\subsection{The rate of spin-flip transitions \label{sec:sf_rate}}

Similarly to the case of localized electrons~\cite{glazov2018electron}, the spin-flip transitions between the bound hole states in sufficiently strong magnetic field where the Zeeman splitting exceeds by far the hyperfine coupling is controlled by the hole-phonon interaction. In moderate magnetic fields studied here, the transitions are mediated by acoustic phonons which can give or take Zeeman energy in the course of spin relaxation. The interaction of Zeeman sublevels with an acoustic phonon is possible since the heavy-hole states $\ket{\Uparrow (\Downarrow)}$ have an admixture of light holes in the presence of magnetic field, as follows from Eq.~\eqref{eq:wfs_mixed}. Hence, these states are coupled under phonon-induced deformation through the Bir-Pikus Hamiltonian~\cite{ref:Woods2004hsr}. Here the complex valence band structure facilitates direct spin-phonon interaction.

The spin-flip transition rates are found using Fermi's golden rule, e.g. the rate of a $\ket{\Uparrow} \to \ket{\Downarrow}$ transition with emission of a phonon is
\begin{equation}
    \label{sf_rate}
    \Gamma_{\Downarrow \Uparrow} = \frac{2 \pi}{\hbar} \sum \limits_{\bm q, \alpha} |M_{\Downarrow \Uparrow}|^2 \delta (\hbar q s_\alpha - |g_{hh}^\perp \mu_B B|)\:,
\end{equation}
where $M_{\Downarrow \Uparrow}$ is the spin-flip matrix element, $\bm q$ is the phonon wave vector, and $s_\alpha$ is the speed of sound in the phonon branch $\alpha$. 
The spin-flip matrix element is $M_{\Downarrow \Uparrow} = \braket{\Downarrow}{\mathcal H_{\rm BP}}{\Uparrow}$, where $\mathcal H_{\rm BP}$ is the Bir-Pikus Hamiltonian~\cite{bir_pikus}, which in the basis~\eqref{eq:newbasis} reads
\begin{multline}
    \label{HBP}
    \mathcal H_{\rm BP} = \left( a' + \frac54 b' \right) \sum \limits_{i} u_{ii} - b' \sum \limits_i F_i^2 u_{ii}
    \\ - \frac{d'}{\sqrt{3}} \sum \limits_{i \neq i'} \left \{ F_{i'} F_i \right \} u_{i' i}\:.
\end{multline}
Here $a'$, $b'$ and $d'$ are the valence-band deformation potentials modified by the Coulomb interaction, $b'/b = d'/d = \int dr r^2 [f^2(r) - 3 g^2(r)/5]$~\cite{Kogan1981}. The phonon-induced deformation results in the strain components 
\begin{equation}
 u_{ij}^{\bm q, \alpha} = \sqrt{\frac{\hbar}{2\rho \omega_{\bm q, \alpha}}} \frac{ \mathrm i \left[ q_i e_j^{(\bm q, \alpha)}+ q_j e_i^{(\bm q, \alpha)} \right]}{2} \mathrm e^{\mathrm i (\bm{q} \cdot \bm{r} - \omega_{\bm q, \alpha} t )} b^\dag_{\bm q, \alpha} + c.c.\:,
 \label{eq:phtensor}
\end{equation}
where $\bm e^{(\bm q, \alpha)}$ is the polarization vector, $\omega$ is the phonon frequency, $\rho$ is the mass density, and $b^\dag_{\bm q, \alpha}$ is the phonon creation operator.
For LA phonons, $\bm e^{(1)} = (q_x,q_y,q_z)/q$, whereas for TA phonons, there are two modes with $\bm e^{(2)} = (q_y,-q_x,0)/q_{\bot}$ and $\bm e^{(3)} = (q_x q_z, q_y q_z,-q_{\bot}^2)/q q_{\bot}$, where $q_{\bot} = \sqrt{q_x^2+q_y^2}$. Further we use the long wavelength approximation for the phonons, i.e. $\mathrm e^{\mathrm i \bm{q} \cdot \bm{r}} \approx 1$.


At zero temperature (when no phonons are present) it follows from Eqs.~\eqref{eq:wfs_mixed}, \eqref{HBP} and \eqref{eq:phtensor} that
\begin{multline}
    \label{M_sf}
    M_{\Downarrow \Uparrow}  = \frac{3 g_0 \mu_B B b'}{2 \Delta_0} \sqrt{\frac{\hbar}{2\rho \omega}} \\
    \times \left( \mathrm i q_x e_z + \mathrm i q_z e_x + q_x e_y + q_y e_x \right)\:,
\end{multline}
where for simplicity we used the spherical approximation $b' = d'/\sqrt{3}$ for the Bir-Pikus Hamiltonian. 
Note that in agreement with time-reversal symmetry, the spin-flip matrix-element is proportional to $B$, so that only magnetic-field induced admixture of light holes in the states~\eqref{eq:wfs_mixed} is relevant for the spin-flip process.
The spin-flip rate calculated after Eqs.~\eqref{sf_rate} and \eqref{M_sf} is
\begin{equation}
    \label{sf_rate_final}
    \Gamma_{\Downarrow \Uparrow} = \frac{\left( g_{hh}^\perp \mu_B B \right )^5}{10 \pi \rho \hbar^4 (u_{xx} - u_{yy})^2} \left( \frac{1}{s_t^5} + \frac{2}{3 s_l^5} \right)\:.
\end{equation}
where we used Eq.~\eqref{g_inplane} to exclude unknown parameter $g_0$ and the relation $\Delta_1 = -b' (u_{xx} - u_{yy})$.

The measured spin relaxation time T$_1$ is related to the spin-flip rate at zero temperature as~\cite{ref:Linpeng2016lsr}:
\begin{equation}
\begin{aligned}
\label{theory:T1:fin}
T_1 = \frac{\mathrm e^\beta - 1}{\Gamma_{\Downarrow \Uparrow}(\mathrm e^\beta + 1)}\:,
\end{aligned}
\end{equation}
where $\beta = |g_{hh}^\perp \mu_B B|/k_B T$, and $k_B T$ is the thermal energy. In the whole range of applied magnetic fields $\beta \ll 1$, and hence, $T_1 = \beta / (2 \Gamma_{\Downarrow \Uparrow}) \propto B^{-4}$. The values of the parameters used to calculate T$_1$ are listed in Table~\ref{table:T1constant}. With $|u_{xx} - u_{yy}| = 0.008 \%$, so that $|u_{xx} - u_{yy}|/|u_{xx}| = 0.2$, the theoretical $T_1$ curve agrees well with the measured data at high fields, shown in Fig.~\ref{fig:T1}(c). It corresponds to $|\Delta_1|/\Delta_0 \approx 0.05$, and using the measured value of $|g_{hh}^\bot| = 0.15$ we estimate $|g_0| \approx 1$. This agrees with the values from literature which have reported hole $g$-factors ranging from 0.52 to 2.34~\cite{ref:Bimberg1978alq,ref:Malyshev1997mma,ref:Kirpichev1996mmc}.

\begin{table}[h]
\begin{center}
\begin{tabular}{c c c c c c c}

\Xhline{2\arrayrulewidth}
$\rho$ (kg/m$^3$) & $s_l$ (m/s) & $s_t$ (m/s) & $|g_{hh}^\bot|$  \\
\Xhline{1\arrayrulewidth}   
$5.32{\times}10^3$\cite{ref:Linpeng2016lsr} & $4.73{\times}10^3$\cite{ref:Linpeng2016lsr}  & $3.35{\times}10^3$\cite{ref:Linpeng2016lsr} & 0.15 \\
\Xhline{2\arrayrulewidth}
\end{tabular}
\end{center}
\caption{Parameters used to calculate T$_1$. Parameter $|g_{hh}^\bot|$ is determined from PL experiments, see Sec.~\ref{sec:strainandPL}}.
\label{table:T1constant}
\end{table}

\section{conlusion}
\label{sec:conclusion}

We have introduced compressive strain into a p-type GaAs epitaxial layer through the epitaxial lift-off technique. This strain breaks the degeneracy of heavy and light hole states, leading to \textmu s-scale longitudinal spin relaxation times T$_1$ of the heavy hole. Coherent population trapping measurements indicate a 7~ns hole spin dephasing time T$_2^*$ in this strained sample. We quantitatively explain the measured T$_1$ and T$_2^*$ values based on two different mechanisms. The measured T$_1$ is explained by a hole-phonon interaction mediated by heavy-hole light-hole mixing in the in-plane magnetic field, and the measured T$_2^*$ is explained by in-plane hole g-factor broadening due to strain inhomogeneity. Other strain engineering techniques such as wafer bonding~\cite{ref:Stanton2020esh} or piezoelectric actuators~\cite{ref:Yuan2018usf} can possibly provide stronger and more homogeneous strain in the sample, which can potentially enhance both T$_1$ and T$_2^*$ in the acceptor-hole system.

\acknowledgements
This material is based upon work supported by the National Science Foundation under Grant Nos. 1150647 and 1820614. M.V.D. acknowledges financial support from the Basis Foundation for the Advancement of Theoretical Physics and Mathematics and the Russian Federation President Grant No. MK- 2943.2019.2. M.M.G. was partially supported by the Saint-Petersburg State University for a research Grant \# 51125686.

\appendix

\section{ELO process}
\label{app:ELO}

In the strained sample, the GaAs epitaxial layer is transferred to the MgO substrate through an ELO technique. The as-grown GaAs sample contains a 2~\textmu m p-type GaAs epitaxial layer with $\sim$50~nm Al$_{0.2}$Ga$_{0.8}$As cladding and an 100~nm AlAs sacrificial layer for ELO, with all those layers on a GaAs substrate. The AlAs sacrificial layer is selectively etched in 5\% hydrofluoric acid for $\sim$24 hours to release the 2~\textmu m GaAs epitaxial layer from the GaAs substrate. A $\sim$1~mm thick photoresist layer is applied to the membrane before the etching for protection. The photoresist needs to be mm thick so the GaAs epitaxial layer can bend with a small angle to let the acid etch in. After the etching, the membrane is transferred to a beaker with water for cleaning and then to an MgO substrate. A thin paper tissue is used to wick the water out from between the epitaxial layer and the MgO substrate. The sample is then put in a SPI membrane box to add pressure on top of the GaAs epitaxial layer to improve bonding with the new substrate. After waiting for $\sim$3 days, the sample is taken out and the photoresist on top of the epitaxial layer is removed in hot solvent.

\begin{figure}[!h]
  \centering
  \includegraphics[width=3 in]{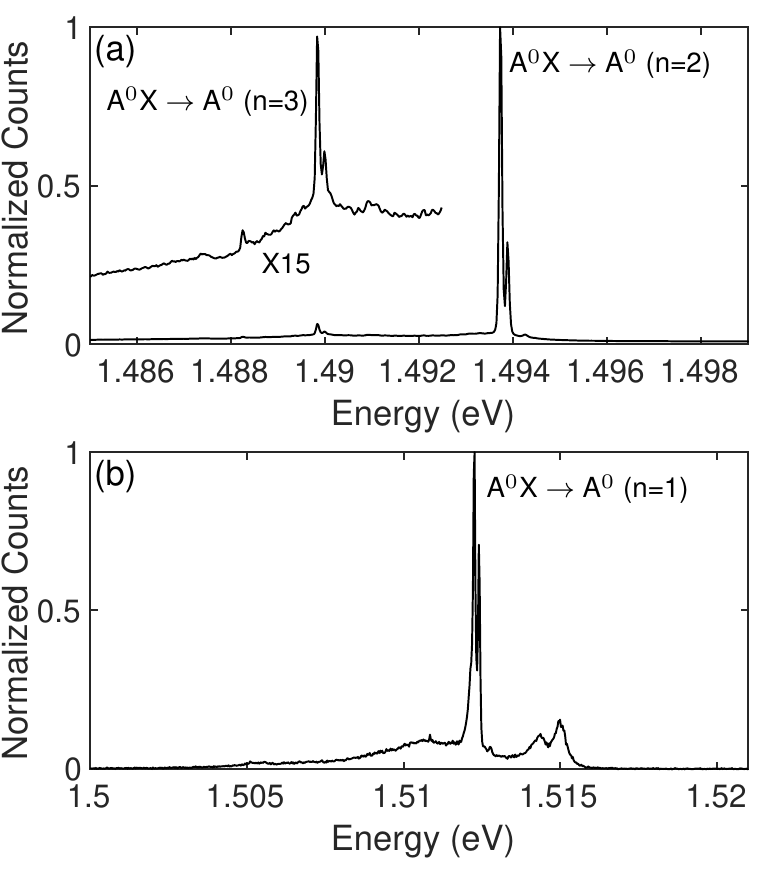}
  \caption{\label{fig:THT} (a) Spectrum of the n=2 and n=3 THT transitions. The laser is resonantly on the main acceptor transition at 1.512~eV with 0.9~\textmu W power. The temperature is at 2~K and the magnetic field is at 0~T. (b) The corresponding spectrum of the main acceptor lines. The laser is at 1.653~eV with 13~nW power. The laser has a spot size of $\sim$1~\textmu m. The spectra are taken in the sample before ELO process.
  }
\end{figure}

\section{Spectrum of THT}
\label{app:THT}

The two hole transitions (THT) are the transitions from the A$^0$X to higher orbital states of A$^0$. The wave function A$^0$ is hydrogen like. The main donor lines are from A$^0$X to n=1 states of A$^0$. The THT transitions are from A$^0$X to n$>$1 states of A$^0$, as shown in Fig.~\ref{fig:THT}. For all the PLE spectra, we collect the signal from the n=2 THT transitions.

\section{In-plane g factors}
\label{sec:gfactor}

\begin{figure}[!htbp ]
  \centering
  \includegraphics[]{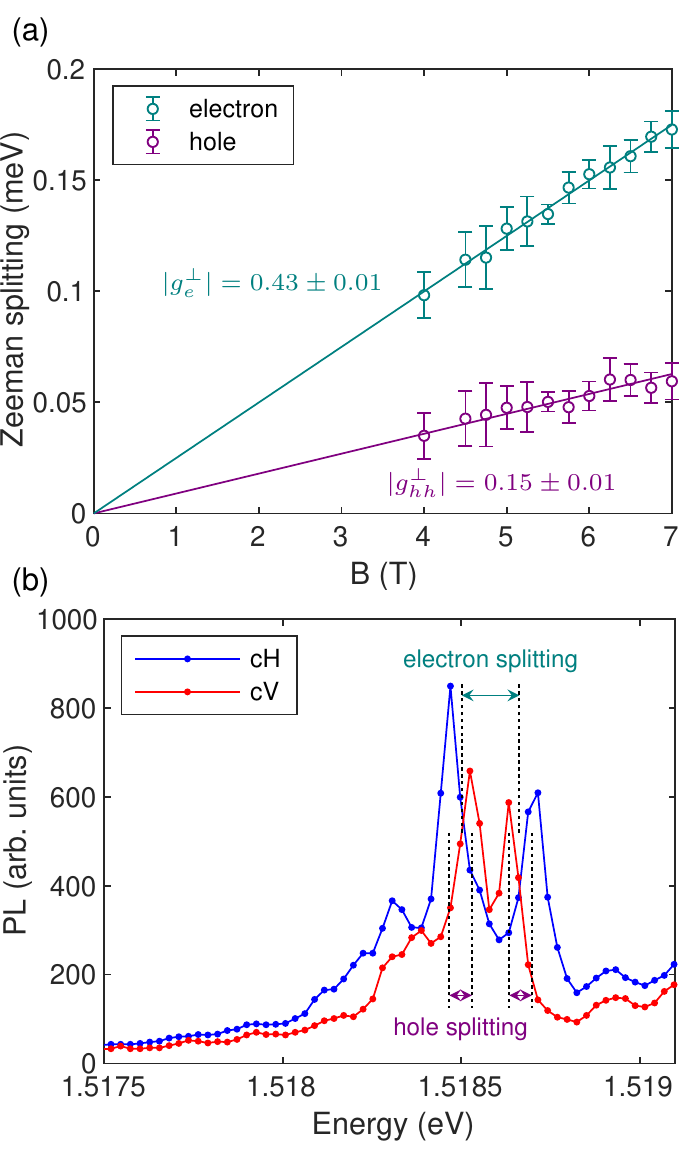}
  \caption{\label{fig:zeemansplitting} (a) Electron and hole Zeeman splitting as a function of the in-plane magnetic field. (b) PL spectra with horizontal and vertical polarization in collection. The magnetic field is at 7~T and the temperature is at 1.5~K. Excitation at 1.53~eV with 200~nW. The electron and hole splittings are marked in the spectra.
  }
\end{figure}

The absolute values of the measured in-plane hole and electron $g$-factors are $|g^{\bot}_{hh}| =  0.15\pm0.01$ and $|g^{\bot}_{e}| = 0.43\pm0.01$, as shown in Fig.~\ref{fig:zeemansplitting}. This values are close to the values measured in InGaAs/GaAs quantum dots~\cite{ref:Tholen2019atg}. The sign of both the electron and hole $g$-factors is negative, as shown in Appendix~\ref{app:selectionrules}.

\section{selection rules}
\label{app:selectionrules}

\begin{figure}[!h]
  \centering
  \includegraphics[width=2 in]{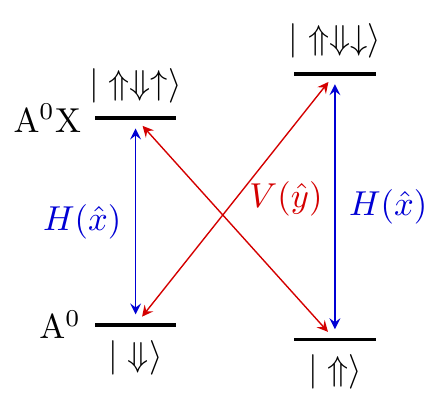}
  \caption{\label{fig:energydiagram} Energy diagram and selection rules of the acceptor system under magnetic fields. $\hat{x}$ is the horizontal direction (parallel to the magnetic field) and $\hat{y}$ is the vertical direction (perpendicular to the magnetic field).
  }
\end{figure}

The selection rules are derived using the same method in Ref.~\cite{ref:Karin2015rpm}. At zero magnetic field, the eigen functions for electrons are
\begin{equation}
\begin{aligned}
\ket{\uparrow}_{B=0} = \ket{\uparrow_z, S}, \\
\ket{\downarrow}_{B=0} = \ket{\downarrow_z, S},
\end{aligned}
\end{equation}
where $\ket{S}$ is the orbital function for electrons. The eigen functions for heavy holes are
\begin{equation}
\begin{aligned}
\ket{\Uparrow}_{B=0} = \ket{\downarrow_z, \frac{X-i Y}{\sqrt{2}}}, \\
\ket{\Downarrow}_{B=0} = \ket{\uparrow_z, \frac{X+i Y}{\sqrt{2}}},
\end{aligned}
\end{equation}
where $\ket{\frac{X\pm iY}{\sqrt{2}}}$ is the orbital function for hole. Note that we use the definition in which $\ket{\Uparrow}_{B=0}$ means missing of a spin up electron so its angular momentum is negative. In an in-plane magnetic field (${B \parallel x}$), the electron states become
\begin{equation}
\begin{aligned}
\ket{\uparrow}_{B>0} = \frac{1}{\sqrt{2}} \ket{\uparrow_z, S} + \frac{1}{\sqrt{2}}\ket{\downarrow_z, S}, \\
\ket{\downarrow}_{B>0} =  \frac{1}{\sqrt{2}} \ket{\uparrow_z, S} -  \frac{1}{\sqrt{2}} \ket{\downarrow_z, S}.
\end{aligned}
\end{equation}
Assuming a non-zero $u_{xx}-u_{yy}$ and a zero $u_{xy}$, the hole states in magnetic fields become 
\begin{equation}
\begin{aligned}
\ket{\Uparrow}_{B>0} = \frac{1}{\sqrt{2}} \left | \downarrow_z, {\frac{X-i Y}{\sqrt{2}}} \right \rangle + \frac{1}{\sqrt{2}} \left | \uparrow_z, \frac{X+i Y}{\sqrt{2}} \right \rangle, \\
\ket{\Downarrow}_{B>0} =   \frac{1}{\sqrt{2}} \left | \downarrow_z, \frac{X-i Y}{\sqrt{2}} \right \rangle - \frac{1}{\sqrt{2}} \left | \uparrow_z, \frac{X+i Y}{\sqrt{2}} \right \rangle.
\end{aligned}
\end{equation}
The dipole matrix element for the recombination of the electron state $|i\rangle$ and hole state $|j\rangle$ is ${\pmb p_{ij}} = \langle j| {\pmb \mu} | i \rangle$. The dipole operator is ${\pmb \mu} = e {\pmb r}$, where $e$ is the electron charge and ${\pmb r} = x  \hat{\pmb x} + y \hat{\pmb y} + z  \hat{\pmb z}$ is the space vector. The calculated results of the four ${\pmb p_{ij}}$ are shown below:
\begin{equation}
\begin{cases}
    \langle \Uparrow | {\pmb \mu} |\uparrow  \rangle = \frac{\mu_0}{\sqrt{2}} \hat{\pmb x} , \\
    \langle \Uparrow | {\pmb \mu} | \downarrow \rangle = -i \frac{\mu_0}{\sqrt{2}} \hat{\pmb y} , \\
    \langle \Downarrow | {\pmb \mu} | \uparrow \rangle = i \frac{\mu_0}{\sqrt{2}} \hat{\pmb y} , \\
    \langle \Downarrow | {\pmb \mu} | \downarrow \rangle = -\frac{\mu_0}{\sqrt{2}} \hat{\pmb x}.
  \end{cases}
  \label{eq:dmatrix2}
\end{equation}
where $\mu_0 = \langle X |e \cdot x | S \rangle = \langle Y |e \cdot y | S \rangle = \langle Z |e \cdot z | S \rangle$. The intensity of the optical transition between electron state $|i\rangle$ and hole state $|j\rangle$ is proportional to $|{\pmb \varepsilon} \cdot {\pmb p_{ij}}|^2$, where ${\pmb \varepsilon}$ is the electric field vector. In this Voigt geometry, $\hat{\pmb x}$ and $\hat{\pmb y}$ are the horizontal and vertical directions. Equation~\eqref{eq:dmatrix2} indicates two transitions are polarized in horizontal direction and two transitions are polarized vertical direction, which matches with our experimental data. Figure~\ref{fig:zeemansplitting}(b) shows that the two transitions with horizontal polarization are at the highest and lowest energy of the four allowed transitions. As the electron $g$-factor is negative, to match with this data, the hole $g$-factor needs to be negative. The corresponding energy diagram of the acceptor system under magnetic fields is shown in Fig.~\ref{fig:energydiagram}. We note that we use the notation that the recombination of $\ket{\Uparrow}$ hole and $\ket{\uparrow}$ electron represents the transition of $\ket{\Uparrow\Downarrow\uparrow} \leftrightarrow \ket{\Downarrow}$.

Assuming a zero $u_{xx}-u_{yy}$ and a non-zero $u_{xy}$, the hole states in magnetic fields become 
\begin{equation}
\begin{aligned}
\ket{\Uparrow}_{B>0} = \frac{1}{\sqrt{2}} \left | \downarrow_z, {\frac{X-i Y}{\sqrt{2}}} \right \rangle + \frac{i}{\sqrt{2}} \left | \uparrow_z, \frac{X+i Y}{\sqrt{2}} \right \rangle, \\
\ket{\Downarrow}_{B>0} =   \frac{1}{\sqrt{2}} \left | \downarrow_z, \frac{X-i Y}{\sqrt{2}} \right \rangle - \frac{i}{\sqrt{2}} \left | \uparrow_z, \frac{X+i Y}{\sqrt{2}} \right \rangle.
\end{aligned}
\end{equation}
The corresponding calculated results of the four ${\pmb p_{ij}}$ are shown below:
\begin{equation}
\begin{cases}
    \langle \Uparrow | {\pmb \mu} |\uparrow  \rangle = \frac{\mu_0}{2} \frac{1-i}{\sqrt{2}} (\hat{\pmb x} - \hat{\pmb y}), \\
    \langle \Uparrow | {\pmb \mu} | \downarrow \rangle = -\frac{\mu_0}{2} \frac{1+i}{\sqrt{2}} (\hat{\pmb x} + \hat{\pmb y}), \\
    \langle \Downarrow | {\pmb \mu} | \uparrow \rangle = \frac{\mu_0}{2} \frac{1+i}{\sqrt{2}} (\hat{\pmb x} + \hat{\pmb y}), \\
    \langle \Downarrow | {\pmb \mu} | \downarrow \rangle = -\frac{\mu_0}{2} \frac{1-i}{\sqrt{2}} (\hat{\pmb x} - \hat{\pmb y}),    
  \end{cases}
  \label{eq:dmatrix1}
\end{equation}
This indicates two transitions are polarized in 45 degree direction ($\hat{\pmb x} + \hat{\pmb y}$) and two transitions are polarized in -45 degree direction ($\hat{\pmb x} - \hat{\pmb y}$). However, this selection rules contradict with our PL measurements shown in Fig.~\ref{fig:zeemansplitting}(b). Therefore, we use a non-zero $u_{xx}-u_{yy}$ and a zero $u_{xy}$ in the calculation of T$_1$.

\section{Density matrix model for CPT}
\label{app:dmCPT}

\begin{figure}[!h]
  \centering
  \includegraphics[width=1.5 in]{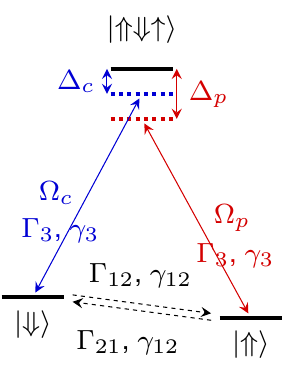}
  \caption{\label{fig:CPTdm} Energy diagram of the $\Lambda$-system.
  }
\end{figure}

In our experiment, the states $\ket{\Uparrow\Downarrow\uparrow}$, $\ket{\Uparrow}$, and $\ket{\Downarrow}$ are used to form the $\Lambda$ system, as shown in Fig.~\ref{fig:CPTdm}. The evolution of the $\Lambda$ system can be simulated by solving the master equation $\partial \rho/ \partial t = -i [H, \rho] + L(\rho)$. In the equation, H is the Hamiltonian of the system
\begin{equation}
    H_i = -\hbar\begin{pmatrix} 
    \Delta_c & 0 & \Omega_c^*/2 \\
    0 & \Delta_p & \Omega_p^*/2 \\
    \Omega_c/2 & \Omega_p/2 & 0 
    \end{pmatrix} \begin{matrix*}[l] 
    \ket{\Downarrow} \\
    \ket{\Uparrow} \\
    \ket{\Uparrow\Downarrow\uparrow}
    \end{matrix*},
\end{equation}
where $\Delta_p$ and $\Delta_c$ is the detuning of the probe and control laser. L is the Lindblad operator including the relaxation and dephasing between different states

\begin{equation}
\begin{aligned}
L & = 
\left(\begin{matrix}
  -\Gamma_{12}\rho_{11}+\Gamma_{21}\rho_{22}+\Gamma_3\rho_{33}\\
  -(\frac{\Gamma_{12}+\Gamma_{21}}{2}+\gamma_{12})\rho_{21} \\
  -(\frac{\Gamma_{12}+2\Gamma_{3}}{2}+\gamma_{3})\rho_{31}
\end{matrix}\right.\\
&\left. \begin{matrix}
-(\frac{\Gamma_{12}+\Gamma_{21}}{2}+\gamma_{12})\rho_{12} & -(\frac{\Gamma_{12}+2\Gamma_{3}}{2}+\gamma_{3})\rho_{13}\\
\Gamma_{12}\rho_{11}-\Gamma_{21}\rho_{22}+\Gamma_3\rho_{33} &
 -(\frac{\Gamma_{21}+2\Gamma_{3}}{2}+\gamma_{s3})\rho_{23}\\
  -(\frac{\Gamma_{21}+2\Gamma_{3}}{2}+\gamma_{3})\rho_{32} & -2\Gamma_{3}\rho_{33}
  
\end{matrix}\right).
\end{aligned}
\end{equation}

$\Gamma_{12}$ and $\Gamma_{21}$ are spin relaxation rate between $\ket{\Uparrow}$ and $\ket{\Downarrow}$. The spin relaxation time $\text{T}_1$ satisfies
\begin{eqnarray}
    \Gamma_{12} = \frac{1}{\text{T}_1} \cdot  \frac{1}{1+e^{-g \mu_B B/k_b T}}, \\
    \Gamma_{21} = \frac{1}{\text{T}_1} \cdot  \frac{e^{-g \mu_B B/k_b T}}{1+e^{-g \mu_B B/k_b T}},
\end{eqnarray}
where $g \mu_B B$ is the hole Zeeman splitting and $k_b T$ is the thermo energy. $\gamma_{12} = 1/\text{T}_2$ is the dephasing rate between $\ket{\Uparrow}$ and $\ket{\Downarrow}$. $\Gamma_3$ and $\gamma_3$ are the spin relaxation rate and dephasing rate between the excited state $\ket{\Uparrow\Downarrow\uparrow}$ and the ground spin states. For simplicity, we assume they are the same for $\ket{\Uparrow}$ and $\ket{\Downarrow}$. $\Delta_c$ and $\Omega_c$ are the detuning and strength of the control laser. $\Delta_p$ and $\Omega_p$ are the detuning and strength of the probe laser. The population of the excited state $\rho_{33}$ after equilibrium is calculated as the final result, which is proportional to the detected PL intensity in the two-laser PLE experiment. All important fitting parameters and the fitting results are shown in Table~\ref{table:CPTfitresult}. 


\begin{table}[!th]
\setlength{\tabcolsep}{0.15 cm}
\begin{center}
\begin{tabular}{c c c c c c c}

\Xhline{2\arrayrulewidth}
 parameter & fit values \\
\Xhline{1\arrayrulewidth}
T$_2^*$~(ns) & $6.8\pm0.7$  \\
T$_1$~(\textmu s) & $0.09\pm0.01$  \\
$\Gamma_3$~(GHz) & $0.63\pm0.03$  \\
$\gamma_3$~(GHz) & $0.64\pm0.05$  \\
$\Omega^2/\text{power}$~(GHz$^2$/\textmu W) & $0.046\pm0.004$ \\
$\Delta_{c0}$~(GHz) & $0.215\pm0.009$  \\
\Xhline{2\arrayrulewidth}
\end{tabular}
\end{center}
\caption{Fitting parameters for the 3-level density matrix model. The errors are 2$\sigma$ errrors from fitting.}
\label{table:CPTfitresult}
\end{table}

\section{Theory of the inhomogeneous dephasing time T$_2^*$ due to hyperfine interaction with nuclear spins}
\label{sec:T2theory}

The hole spin dephasing originates from the dipole-dipole part of the hyperfine interaction. In bulk semiconductors it is given by the Hamiltonian acting on the $4$-component envelope function~\cite{glazov2018electron}
\begin{equation}
    \label{hf:hole}
    \mathcal H_{\rm hf} = \sum \limits_{j, a}\frac{A_a v_0}{2} \delta(\bm r-\bm R_{j,a}) (M_{1,a} \bm I^{j,a}\cdot \bm J + M_{2,a} \bm I^{j,a}\cdot \bm J^3),
\end{equation}
where $a = {\rm Ga,~As}$ is the crystal sublattice index, $j$ enumerates nuclei in a given sublattice, $v_0$ is the  volume of the unit cell, $\bm R_{j,a}$ is the  position of $j$th nucleus in the sublattice $a$, $A_a$ is the conduction band hyperfine coupling constant, $\bm I^{j,a} = (I_x^{j,a},I_y^{j,a},I_z^{j,a})$ is the spin of the nucleus, and we recall that $\bm J$ is the free-hole angular momentum and we use the notation $\bm J^3 =(J_x^3,J_y^3,J_z^3)$. In Eq.~\eqref{hf:hole} $M_{1,a}$ and $M_{2,a}$ are the dimensionless parameters with $M_{2,a}$ resulting from the cubic symmetry of the crystal. Depending on the material system and isotope in question the parameters $M_{1,a}$ and $M_{2,a}$ can be comparable~\cite{glazov2018electron,Chekhovich:2013ys,PhysRevB.94.121302}. Hereafter, for simplicity, we take into account the contribution of the $M_{1,a}$ only: As we see below it already produces  right order-of-magnitude of the dephasing time.

In our experiments, the hole spin dephasing is studied in the Voigt configuration where $\bm B \parallel x$. In the external field along the $x$-axis, the hole-states $\ket{\Uparrow}$ and $\ket{\Downarrow}$ are given by Eq.~\eqref{eq:wfs_mixed} with the $\ket{\pm 3/2}$ and $\ket{\pm 1/2}$ basis functions given by the general expressions~\eqref{eq:newbasis}.  Provided that the field is sufficiently strong, i.e., under the conditions where the hole Zeeman splitting in the magnetic field, $|g_{hh}^\perp \mu_B B|$, Eq.~\eqref{g_inplane}, exceeds by far the splittings induced by the nuclear spin fluctuations, the dephasing is controlled by  the nuclear field fluctuations in the directions of the hole pseudospin. We calculate  $\Omega_{N,x}$, the contribution to Larmor frequency due to nuclei, as
\begin{equation}
\hbar \Omega_{N,x} = \braket{\Uparrow}{\mathcal H_{\rm hf}}{\Uparrow} - \braket{\Downarrow}{\mathcal H_{\rm hf}}{\Downarrow}.
\end{equation}
The analysis shows that it contains $B$-independent and $B^2$ terms, the latter are neglected. Correspondingly, the mean square fluctuation of nuclear field is
\begin{multline}
\aver{\Omega_{N,x}^2} = \frac{v_0^2 \sum_a C_a^2 I_a(I_a+1)}{27 \hbar^2}  \\
\times \sum \limits_{j} \left[ j_x^2(\bm R_j) + j_y^2(\bm R_j) + j_z^2(\bm R_j) \right]\:,
\end{multline}
where $C_{a} = 3 A_a M_{1,a}/2$, and  $j_\alpha (\bm r) = \braket{\Uparrow}{J_\alpha}{\Uparrow} - \braket{\Downarrow}{J_\alpha}{\Downarrow}$. Here we took into account that $\aver{I_\alpha I_\beta} = \delta_{\alpha \beta} I(I+1)/3$.
Summation by $j$ can be changed to the integration using the standard expression
\begin{equation}
\int F(\bm R) d \bm R = v_0 \sum \limits_j F(\bm R_j)\:.
\end{equation}
Using this formula and taking into account that nuclei in GaAs have the same spin $I_a = I = 3/2$, we finally obtain
\begin{multline}
\aver{\Omega_{N,x}^2} = \frac{v_0 I(I+1) (C_{\rm As}^2 + C_{\rm Ga}^2)}{27 \hbar^2} \\
\times \int d\bm r \left[ j_x^2(\bm r) + j_y^2(\bm r) + j_z^2(\bm r) \right]\:. 
\end{multline}

The compact analytical expression for the $\langle\Omega_{N,x}^2\rangle$ can be derived assuming that $|f(r)|$ exceeds by far $|g(r)|$ in Eqs.~\eqref{eq:newbasis}. Keeping the terms with the lowest powers of $g(r)$ we arrive at
\begin{multline}
\label{omega}
\aver{\Omega_{N,x}^2} = \frac{v_0 I(I+1) (C_{\rm As}^2 + C_{\rm Ga}^2)}{9 \pi \hbar^2} \\
\times \int d r  r^2 \left[ \frac{2 f(r)^2 g(r)^2}{5} + \frac{3 \Delta_1^2 f(r)^4}{4 \Delta_0^2} \right] \:.
\end{multline}
It is noteworthy that Eq.~\eqref{omega} contains two contributions. The second one $\propto (\Delta_1/\Delta_0)^2$ is related to the heavy-light hole mixing due to the strain and is similar to the one widely studied in quantum dot structures~\cite{ref:Testelin2009hsd,glazov2018electron}. The first contribution results from the complex structure of the acceptor function. This contribution does not require any strain.

Assuming the Gaussian distribution of the nuclear field fluctuations~\cite{ref:Merkulov2002esr}, the component of the hole pseudospin normal to magnetic field $\bm S_\perp$ decays as $\bm S_\perp \sim \exp(-\aver{\Omega_{N,x}^2} t^2/2)$, the corresponding decay time $T_2^* = \sqrt{2} \aver{\Omega_{N,x}^2}^{-1/2}$. Using Eq.~\eqref{omega}, $C_{\rm As} = 4.4~\mu$eV, $C_{\rm Ga} = 3~\mu$eV~\cite{ref:Testelin2009hsd}, $\Delta_1/\Delta_0 = 0.05$, $\int dr r^2 f^2 g^2 = 0.5/a_B^3$, $\int dr r^2 f^4 = 7.9/a_B^3$, where $a_B$ is the acceptor Bohr radius~\cite{ref:Baldereschi1973sms}, we obtain $T_2^* \approx 58$~ns for the first contribution in Eq.~\eqref{omega} and $T_2^* \approx 216$~ns for the second contribution in Eq.~\eqref{omega}.

\bibliography{xiayu.bib}

\end{document}